\documentclass[aps,prd,onecolumn,nofootinbib,superscriptaddress]{revtex4-2}
\usepackage[T1]{fontenc}
\usepackage[utf8]{inputenc}
\usepackage{amsmath,amssymb,mathtools,bm}
\usepackage{graphicx}
\usepackage{hyperref}
\makeatletter
\let\origshowhyphens\showhyphens
\makeatother

\usepackage{microtype}

\makeatletter
\let\showhyphens\origshowhyphens
\makeatother
\usepackage{physics}
\usepackage{enumitem}
\usepackage{tikz}
\usepackage{newtxtext,newtxmath}
\usepackage{pgfplots}
\pgfplotsset{compat=1.18}

\pgfplotsset{
    every axis/.style={
        line width=0.6pt,
        tick style={line width=0.5pt},
        font=\footnotesize,
        label style={font=\footnotesize},
        tick label style={font=\footnotesize},
    }
}
\hypersetup{colorlinks=true,linkcolor=blue,citecolor=blue,urlcolor=blue}

\begin{document}

\title{Gravitational Redshift of Light and the Heisenberg Uncertainty Principle}

\author{Asher Klatchko}
\affiliation{Portland, Oregon}
\email{klatchko@reed.edu}

\author{Robert Hill}
\affiliation{Los Alamos National Laboratory, New Mexico}
\email{robert.hill@lanl.gov}

\begin{abstract}
Empirical observations together with theoretical analyses are being used to argue that the classical phenomenon of gravitational redshift---namely, the redshift of light in a static gravitational potential---may be in tension with the Heisenberg uncertainty principle. In particular, in the Pound--Rebka experiment, the emitter--absorber pair interact with each other by exchanging a M\"ossbauer photon, a process that is fully describable within quantum mechanics but how the photon interacts with the spacetime metric remains unclear. Since the uncertainty principle is incompatible with local realism, as per general relativity, we propose to study continuous-variable photonic entanglement within the Einstein--Podolsky--Rosen (EPR) framework in a weak gravitational field. We outline a thought experiment, realizable on the surface of the Earth, that could shed further light on this problematic seamline between general relativity and quantum mechanics in the weak-field regime.
\end{abstract}

\maketitle

\noindent\textbf{Keywords:} Gravitational redshift of light, Heisenberg uncertainty, photonic entanglement vis-\`a-vis EPR

\section{Introduction}

Gravitational redshift (GRS) can be derived from first principles and was predicted by Einstein, on the basis of the equivalence principle, prior to the formulation of general relativity (GR) \cite{Einstein1911}. Subsequent analyses have argued that, unless the energy of a photon traversing a static gravitational potential decreases, the first law of thermodynamics would be violated\footnote{The argument proceeds as follows. Consider an excited atom at point 2 with energy, $E_{2}=m_{A}c^{2}+\epsilon = m_{A}^{*}c^{2}$, falling in a gravitational field to a point 1. Gravity does work, $m_{A}^{*}\Delta\Phi$. Let the atom radiate a photon at point 1, $\epsilon = h\nu_{1}$, thereby reducing the atom's energy to $m_{A}c^{2}$. Now the atom is taken back to point 2. Gravity does work, $-m_{A}\Delta\Phi$. If the emitted photon's frequency is not changed it could excite an identical atom causing a cyclic process in which work is done and energy is produced, $\oint dW = \left(m_{A}^{*}-m_{A}\right)\Delta\Phi = \epsilon\frac{\Delta\Phi}{c^{2}}$. Unless the energy of the photon is decreased by GRS such that $h(\nu_{1}-\nu_{2})=\epsilon\frac{\Delta\Phi}{c^{2}}$, the first law of thermodynamics would be violated \cite{Walecka2007}. } \cite{Walecka2007}. Note however that in the covariant formalism of GR, the actual conservation law in any static spacetime with an independent metric 
satisfying $g_{0j} = 0$ is,
\[
E=-p_{0}=-g_{00}p^{0}=|g_{00}|^{1/2} \, E_{\text{local}} = \text{constant}.
\]
This expresses the law of energy redshift, describing how the locally measured 
energy of any particle---massless or massive---changes as it 
"climbs out" of or "falls into" a static gravitational field. With $E_{\text{local}} = h~c/\lambda_{\text{local}}$, this conservation law becomes\footnote{See for example Equation (8) below. }:
\[
|g_{00}|^{-1/2} \, \lambda_{\text{local}} = \text{constant}.
\]
Thus, it is not the local energy itself that is conserved between different observers, but rather the above combination --- the energy at infinity in terms of these locally measured quantities \cite{MTW}. Hence, for a photon moving freely from a given observer to infinity, the correction factor reduces to unity and the local energy (as measured at infinity) becomes identical with the constant E. 

As a quantum object, any change in the photon's energy and momentum must comply with the Heisenberg uncertainty principle (HUP). However, cross-verification of the momentum--position uncertainty relation against empirical data reveals a persistent tension. Reformulating this relation solely in terms of wavelength leads to paradoxical constraints that conflict with GR. Furthermore, because photon momentum is not conserved, we develop a model from first-principles that independently substantiates these inconsistencies. Taken together, these results indicate an incompatibility between GR and QM that manifests itself even within the weak-field regime.

\section{Tension between General Relativity and Quantum Mechanics}

In GR, time dilation causes a clock based on an atomic or molecular transition at ground level to tick differently than one held aloft at height $z$ above the ground. Accordingly, the frequency of a photon at altitude $z$ is related to that at ground level by:
\begin{equation}
\Delta\nu_{z} = \nu_{z} - \nu_{0}
= \frac{\nu_{0}gz}{c^{2}\left(1+\frac{z}{R_{E}}\right)}
\approx \nu_{0}z\times\left(1.09\times10^{-16}\right).
\tag{1}
\end{equation}
Where $R_{E}$ is Earth's radius and $z$ is the altitude measured in meters \cite{PR}. Equation (1) can be generalized to describe the frequency shift between any two altitudes $z$ and $z'$, on Earth's surface provided that $z',z\ll R_{E}$,
\begin{equation}
\Delta\nu_{z'z} = \left(\nu_{z'}-\nu_{z}\right)\approx \nu_{0}(z'-z)\times 1.09\times10^{-16} ~ \rightarrow ~ \frac{\Delta\nu_{z'z}}{\nu_{0}} \approx \Delta_{zz'}\times1.09\times10^{-16}. 
\tag{2}
\end{equation}

\subsection{GRS and the Momentum--Position uncertainty}

In QM, the momentum of a photon is given by:
\[
p=\frac{h}{\lambda},
\]
so the change in wavelength implied by Equation (1) corresponds directly to a change in the photon's momentum. Moreover, the expectation values of the photon's momentum and position become interdependent and are constrained by the HUP. For a photon propagating in the $z$-direction above ground, the uncertainties in position and momentum satisfy:
\[
\Delta p_{z}\Delta z \ge \frac{\hbar}{2},
\]
where $\Delta z$ denotes the uncertainty in the photon's position along the propagation axis and $\Delta p_{z}$ the corresponding uncertainty in its momentum.

We can recast these uncertainties in terms of the photon wavelength as following:
\begin{subequations}
\begin{align}
\Delta p &= \sqrt{\left(\delta\lambda\times\frac{dp}{d\lambda}\right)^{2}} = h\frac{\delta\lambda}{\lambda^{2}},
\tag{3a}\\
\Delta z &= \kappa\lambda.
\tag{3b}
\end{align}
\end{subequations}
where $\kappa>0$ is a dimensionless parameter that may depend on the resolution, $\left(\delta\lambda/\lambda\right)$, of the wavelength (e.g. the natural bandwidth at FWHM). Categorically, any characteristic length scale can be expressed in units of the photon's wavelength. Examples include the coherence length $L_{c}\approx \lambda/(\delta\lambda/\lambda)$, where $\delta\lambda/\lambda$ denotes the spectral resolution; and the beam waist $w_{0}=\lambda/(\pi\theta)$, with $\theta$ the beam's divergence angle. If the photon is modeled as a wavepacket of spatial width $\Delta z=\sigma$, and thereby a minimum-uncertainty state, a monochromatic wavepacket cannot exhibit a spread smaller than its wavelength $\lambda$. Consequently, expressing the positional uncertainty as $\Delta z=\kappa\lambda$ requires $\kappa\ge 1$. Generally, the positional uncertainty may be expressed as a function of the spectral resolution $\delta\lambda/\lambda$ and the photon's wavelength:
\[
\Delta z = \lambda f\left(\frac{\delta\lambda}{\lambda}\right).
\]
Expanding $f$ in a Taylor series about $\delta\lambda/\lambda\to 0$, we obtain:
\[
\Delta z = \lambda\left[f(0)+f'(0)\left(\frac{\delta\lambda}{\lambda}\right)+f''(0)\left(\frac{\delta\lambda}{\lambda}\right)^{2}+\ldots\right].
\]
To leading order this reduces to Equation (3b)\footnote{Consider the following, we can recast the position in terms of the wavelength as, $z=q\lambda$, where $q$ is the number of wavelengths that fit into the distance $z$. For one wavelength we have, $\delta z_{1}\sim\delta\lambda \equiv (\delta\lambda/\lambda)\lambda$, with $\delta\lambda/\lambda$, the resolution of the wavelength, usually related to the FWHM. The RMS of these $q$ deviations is, $\Delta z_{\mathrm{rms}}=\sqrt{\sum \frac{1}{q}\left(\frac{\delta\lambda_{i}}{\lambda}\right)^{2}\lambda^{2}}=\left(\frac{\delta\lambda}{\lambda}\right)\lambda$, with a maximum uncertainty, $\Delta z_{\max}=\sqrt{q\left(\frac{\delta\lambda}{\lambda}\right)^{2}\lambda^{2}}=\sqrt{q}\left(\frac{\delta\lambda}{\lambda}\right)\lambda$. The resolution is a numerical factor and the numerical value is renamed as, $\sqrt{q}\left(\frac{\delta\lambda}{\lambda}\right)\equiv\kappa$, leading to $\Delta z=\kappa\lambda$.}
\[
\Delta z \simeq \lambda f(0)=\kappa\lambda.
\]

Combining (3a) \& (3b) we end up with:
\begin{subequations}
\begin{align}
\eta^{\mathrm{HUP}} &= \Delta p\times\Delta z = h\kappa(\delta\lambda/\lambda)\ge \frac{h}{2\pi},
\tag{4a}\\
\frac{1}{2\kappa\pi} &\le (\delta\lambda/\lambda).
\tag{4b}
\end{align}
\end{subequations}
We note that, by Liouville’s theorem, $\eta^{\mathrm{HUP}}$—representing the volume occupied by the photon’s possible positions and momenta in phase space—must remain conserved and therefore cannot deviate, either upward or downward, from its initial minimum value,$\frac{h}{2\pi}$, during propagation \cite{VanVleck}. We further note that the Planck constant, $h$, drops out of Equations (4), underscoring their intrinsically classical character. 

The wavelength shift of a photon ascending from Earth's surface to an height $z$ is calculated in GR with the Schwarzschild metric \cite{LanLif1998}:
\begin{equation}
\frac{\Delta\lambda}{\lambda}\approx \frac{2M_{E}G}{c^{2}R_{E}^{2}}=\frac{gz}{c^{2}}.
\tag{5a}
\end{equation}
Also note that Equation (5a) can be reformulated as:
\begin{equation}
h\frac{\Delta\lambda}{\lambda}=h\frac{gz}{c^{2}}.
\tag{5b}
\end{equation}

Evidently, Equations (5) presuppose a well-defined value for the photon's momentum $h/\lambda$ along with a definite position $z$. However, the HUP forbids the simultaneous assignment of sharp values to these conjugate variables. We will demonstrate that the empirical data from the Pound and Rebka experiment \cite{PR} are in tension with Equations (4). The Pound--Rebka experiment employed a M\"ossbauer gamma-ray source with an energy of $14.4\,\mathrm{keV}$, corresponding to a wavelength of approximately $\lambda\approx 8.63\times10^{-11}\,\mathrm{m}$. The full width at half maximum (FWHM) of the spectral line was about $4.7\times10^{-9}\,\mathrm{eV}$ \cite{PR}. This yields a spectral resolution of $(\delta\lambda/\lambda)\approx 0.32\times10^{-12}$, whereas the gravitational redshift (GRS) measured in the experiment was $\Delta\lambda/\lambda\approx 2.5\times10^{-15}$. For Equation (4b) to hold true we would need a value of:
\begin{equation}
\kappa \ge (2\pi\delta\lambda/\lambda)^{-1}\approx 2.5\times10^{11},
\tag{6}
\end{equation}
which deems the estimated uncertainty in position for a wavelength of $\lambda\approx 8.63\times10^{-11}\,\mathrm{m}$ to be:
\begin{equation}
\Delta z=\kappa\lambda \ge 21.5\,\mathrm{m}.
\tag{7}
\end{equation}
This value is comparable to the height of the tower ($z=22.5\,\mathrm{m}$) in the Jefferson Laboratory at Harvard University, where the experiment was conducted. This does not contradict the coherence length of a M\"ossbauer photon, $L_{c}\approx \lambda/(\delta\lambda/\lambda)\sim 260\,\mathrm{m}$, which simply reflects the probability of finding the photon over that distance. The problem, rather, is how to conceptually reconcile the uncertainty---or, for that matter, the coherence length---with the strong height dependence implied by Equation (1).

Quantum theory describes the Mössbauer emission mechanism as a single quanta:

\[
|1\rangle = \int d\omega\, f(\omega)\, |1_\omega\rangle,
\]

with a Lorentzian distribution around \(\omega_0\), and a width \(\Gamma\),

\[
f(\omega) \propto [(\omega - \omega_0) + i\Gamma/2]^{-1}:
\]

As the linewidth is extremely narrow, on the order of \(10^{-9}\,\text{eV}\), the emitted pulse has a correspondingly long lifetime of about \(141\,\text{ns}\). According to the Weisskopf–Wigner theory, the photon inherits the exponential decay profile of the nuclear transition, leading to a time-domain wavefunction of the form \cite{ScuZub}:

\[
\psi(t) \propto e^{-i\omega_0 t}\, e^{-t/(2\tau)}\, \Theta(t).
\]

Here, $\Theta(t)$ represents the sharp onset of the emission, while the decay is purely exponential. As a result, the photon is emitted over a characteristic time $\tau$, so its wavepacket length is $L\sim c\tau\approx 40\,\mathrm{m}$. This is smaller than the coherence length $L_{c}$ by a factor of $2\pi$, reflecting the ratio $h/\hbar$. Obviously, this description is not local---neither in space nor in time---since the leading edge of the wavepacket experiences substantially more time dilation than its tail. It may be tempting to imagine that a wavepacket ``samples'' multiple spacetime regions, each with its own redshift, or that its leading and trailing components persist in quantum superposition. But within a QFT framework, such a picture is untenable. Any putative component of the wavepacket that experiences a distinct gravitational redshift must interact locally with the corresponding region of the background geometry. This immediately elevates those supposed ``components'' to the status of independent physical degrees of freedom, not mere substructures of a single coherent excitation. The unavoidable consequence is that gravitational time dilation forces different segments of the wavepacket to accumulate inequivalent phases, and this phase dispersion is flatly incompatible with the preservation of coherence \footnote{Covariantly: in a static spacetime with timelike Killing field $\xi^\mu=(\partial_t)^\mu$, a photon with wavevector $k^\mu$ has conserved
$E_\infty=-k_\mu\xi^\mu=-k_0$.  A static observer at height $z$ has
$u^\mu(z)=\xi^\mu/\sqrt{-g_{00}(z)}$, so the locally measured frequency is
$\omega(z)=-k_\mu u^\mu = E_\infty/(\hbar\sqrt{-g_{00}(z)})$.  In the weak--field
limit $g_{00}(z)\simeq -(1+2\phi(z)/c^{2})$ with $\phi(z)=gz$, one finds
$\omega(z)\simeq \omega_{0}(1-gz/c^{2})$, $\omega_{0}=E_\infty/\hbar$.  The
proper--time phase is $\phi(\tau;z)=-\omega(z)\tau$, so for an extended packet
of length $L$ the proper--time phase difference between $z_{1}$ and
$z_{2}=z_{1}+L$ is
$\Delta\phi(\tau)\simeq \omega_{0}\tau\,gL/c^{2}$}. 
%%%%%
\iffalse
This can be approximated as follows. Consider a photon wavepacket of coherence length

\[
L \sim c\tau ,
\]

propagating vertically in a weak, static gravitational field with Newtonian
potential $\phi(z) = gz$.  In the weak--field limit the metric component is

\[
g_{00}(z) \simeq -\left(1 + \frac{2\phi(z)}{c^{2}}\right),
\]

so the proper time at height $z$ satisfies

\[
d\tau(z) \simeq \left(1 + \frac{\phi(z)}{c^{2}}\right) dt.
\]

A monochromatic mode of coordinate frequency $\omega_{0}$ therefore acquires a
local proper frequency

\[
\omega(z) = \omega_{0}\left(1 + \frac{\phi(z)}{c^{2}}\right)
          = \omega_{0}\left(1 + \frac{gz}{c^{2}}\right).
\]

Hence the phase of the field at position $z$ and coordinate time $t$ is

\[
\phi(t,z) = -\omega(z)\, t
          = -\omega_{0}\left(1 + \frac{gz}{c^{2}}\right)t.
\]

For an extended wavepacket, different points along its length occupy different
heights.  Let the front of the packet be at $z_{1}$ and the tail at
$z_{2} = z_{1} + L$.  The phase difference between these two points at fixed
coordinate time $t$ is

\[
\Delta\phi(t)
= \phi(t,z_{2}) - \phi(t,z_{1})
= -\omega_{0} t\,\frac{gL}{c^{2}}.
\]

\fi
%%%%%%%
Thus the packet develops a linear phase gradient along its spatial extent:

\[
\phi(t,z) = -\omega_{0} t
            -\omega_{0} t\,\frac{gz}{c^{2}},
\]

so that each slice of the wavepacket oscillates with a slightly different
proper frequency.  Over long coherence times, this produces an accumulated
dephasing between the leading and trailing edges proportional to $gL/c^{2}$.

One may reasonably argue that the measurement protocol in the Pound--Rebka experiment \cite{PR} is correctly described by an absorption-probability functional that does not violate the Heisenberg uncertainty principle. For example, consider the description of the fractional counting rate in the experiment as a Lorentzian probability distribution, where the linewidth $\Gamma$ enters as the width of the Lorentzian while the gravitational redshift produces a deterministic shift of its center, and the Doppler-compensating velocity, $v$, is used to restore resonance:
\[
R(v)\propto \frac{\Gamma^{2}}{(h\nu\,v/c-\Delta E_{g})^{2}+\Gamma^{2}}.
\]
For this to hold, the absorber must resonate with the emitted photon, such that the absorber's probability effectively selects the portion of the photon's spectrum that overlaps with its resonance:
\[
P_{\mathrm{abs}}\propto \left|\int d\omega\,A_{\mathrm{em}}(\omega)A_{\mathrm{abs}}^{*}(\omega)\right|^{2}.
\]
According to quantum mechanics, the emitter--absorber pair interacts by exchanging a M\"ossbauer photon, yet how this photon couples to the spacetime metric---specifically in a way that leads to an energy loss---remains obscure. In the quantum picture, the photon mediates a coherent, recoil-free exchange with spectral properties fixed by the transition dynamics, whereas general relativity treats it as a classical test particle whose frequency is deterministically shifted by the gravitational potential via time dilation. Reconciling these descriptions is nontrivial. If the photon's energy--momentum is not well defined during flight but becomes definite only upon measurement, the cumulative redshift predicted by GR becomes conceptually puzzling. Conversely, if the gravitational field acts continuously on the photon, this presupposes a quasi-classical trajectory with well-defined dynamical variables, in tension with the Heisenberg uncertainty principle. This reveals a deeper gap in our understanding of how quantum systems couple to classical spacetime, even in weak fields.

The observed agreement with the GR prediction, $\Delta\nu_{\mathrm{obs}}/\Delta\nu_{\mathrm{GR}}\simeq 1$, only sharpens the issue: according to Equation (1), each additional meter of elevation should dramatically degrade this accuracy. What, then, is the operational meaning of the uncertainty principle for a photon wavepacket propagating in Earth's gravitational field, and how does this comport with the deterministic structure of general relativity? If the photon's momentum is not a real, well-defined property across its entire wavepacket during its ascent, how can the cumulative effect implied by Equation (5) arise? In this view, spacetime effectively acts as a continuous probe of the photon's spatial and temporal properties---an assumption that implicitly reinstates a form of local realism. The fact that the measurement procedure respects the uncertainty principle does not resolve the paradox: gravitational redshift appears to rely on a quasi-classical, locally realistic account of the photon's dynamical variables, in direct tension with the quantum-mechanical prohibition against their simultaneous definiteness.

\subsection{Events vs.\ Photons}

Examination of the notion of a quantum event as the bridge between the virtual character of quantum states and the concrete phenomena observed in space and time is given by Haag in Ref.~\cite{Haag}. Haag argues that individual particles are neither real nor localized in the traditional sense, whereas events possess both reality and locality. An “event’’ provides the operational link between the abstract quantum state and phenomena that are documentable and unquestionably real. Reality and locality are therefore attributed to events rather than to particles. A particle does not carry an inherent position; a measured value is created by the interaction process and is thus an attribute of an event. Individual events have an approximate spacetime location, but a point in space–time cannot be defined by the position of a single particle and must instead be approached operationally through an event.

The Pound--Rebka experiment proceeded via counts of single detected events. A ``count'' (or the detection of an event) was registered by a detector placed behind the absorber\footnote{An NaI scintillation counter was positioned behind the absorber to detect the transmitted $\gamma$ rays.}. 
A higher count rate indicated higher absorption, showing that the Doppler shift was successfully matching the gravitational redshift. 
A transducer oscillating at $74\,\text{Hz}$ was used to move the $^{57}\text{Fe}$ source together with an offsetting velocity of $0.7\,\mu/s$ for modulation, in order to maximize absorption. This corresponds to an oscillatory amplitude of approximately $1.75\,\mu\text{m}$ on top of the continuous drift. The system allows to compensate the mismatch in frequencies between the emitter and the absorber due to gravitational redshift and at the same time to scan for an absorption event. In Haag's language the experiment is not tracking the photon's energy from emission till absorption, but rather enforces a condition on a single coherent phase linking the two events. The redshift is merely a failure of this phase to match when projected onto two different local clocks. The Doppler motion restores the global phase consistency between the two events. Coherence is revealed through measurements between observable events. 

In Haag’s ontology, the burden is shifted away from photon propagation and placed instead on the notion of coherence (see appendix ~\ref{app:haag_phase}), while emphasizing that decoherence alone is insufficient without a further principle of random realization selecting one of the possibilities allowed by the quantum state. Decoherence suppresses interference but does not determine a unique outcome, and Haag argues that an additional principle is required for the emergence of definite events. He invokes the “Heisenberg cut,” originally introduced by Heisenberg as a freely chosen boundary separating quantum superposition dynamics from classical recorded outcomes. Strictly speaking, this arbitrary cut sidesteps—as far as the photon’s position and momentum are concerned—the tension noted above, yet the question of how coherence is empirically established arises. We return to this issue in the sections ahead.
  
\subsection{GR Imposed Constraints on $\kappa$}

The gravitational redshift of a photon emitted from the surface of a massive body at radius $r_{E}$ and observed at earth at a distance $r_{O}\gg r_{E}$, is calculated by the Schwarzschild metric as follows:
\begin{equation}
\frac{\lambda_{O}}{\lambda_{E}}
=
\frac{\sqrt{1-\frac{2GM}{r_{O}c^{2}}}}{\sqrt{1-\frac{2GM}{r_{E}c^{2}}}}
~~r_{0}\to\infty~~
\frac{1}{\sqrt{1-\frac{2GM}{r_{E}c^{2}}}}.
\tag{8}
\end{equation}
The quantity $\epsilon=\frac{2GM}{r_{E}c^{2}}$ is the ratio between the Schwarzschild radius, $\frac{2GM}{c^{2}}$, and the physical radius of the star, $r_{E}$. The fractional shift is, therefore:
\begin{equation}
\frac{\left(\lambda_{O}-\lambda_{E}\right)}{\lambda_{E}}
=
\frac{\Delta\lambda}{\lambda}
=
\frac{1}{\sqrt{1-\epsilon}}-1.
\tag{9}
\end{equation}
GRS values increase as $\epsilon$ approaches the size of its black hole radius. Indeed, as $\epsilon\rightarrow 1$, the shift becomes infinite. An infinite shift is consistent with the fact that light cannot escape the horizon of a black hole.

In what follows, we assume that the numerical values predicted by Eq.~(9) cannot be smaller than the signal detected by a spectrometer’s correlator, which can only broaden the fractional shift $\Delta\lambda/\lambda$, as no detector can reduce the intrinsic spread of the state itself, originating from the operator algebra of QM, $[\hat{x},\hat{p}] = i\hbar$. Detectors typically increase the effective uncertainty through decoherence rather than decrease it. 
\\
\\
%%%%%%%%%%%FIGURE 1
\begin{figure}[t] %!ht
\centering
\begin{tikzpicture}
%\centering
% ================= (a) =================
\begin{axis}[
    name=left,
    width=8.6cm,
    height=5.2cm,
    title={Theoretical Gravitational Redshift},
    xlabel={$\epsilon = \frac{2GM}{Rc^2}$},
    ylabel={$\Delta\lambda/\lambda$},
    ylabel style={
    at={(axis description cs:0.0,1.0)},
    anchor=south,
    rotate=270
    },
    ytick={0.05,0.10,0.15,0.20},
    scaled y ticks=false,
    yticklabel style={
    /pgf/number format/fixed,
    /pgf/number format/precision=2
    },
    scaled x ticks=false,
    xticklabel style={
    /pgf/number format/fixed,
    /pgf/number format/precision=2
    },
    xmin=0, xmax=0.25,
    ymin=0, ymax=0.16,
    axis lines=left,
    domain=0:0.25,
    samples=200,
]

% Redshift curve
\addplot[black, thick] {sqrt(1/(1-x)) - 1};

% Reference line % Vertical marker at ε = 0.142
\addplot[dashed] coordinates {(0, {1/(4*pi)}) (0.142, {1/(4*pi)})};
\addplot[dashed] coordinates {(0.142,0) (0.142,1/(4*pi)};
\node[left] at (axis cs:0, {1/(4*pi)}) {$(4\pi)^{-1}$};
\coordinate (yposA) at (axis cs:0,0.0796);
\node at (axis cs:0.1,0.09) {\small $\kappa = 1$};
\node at (axis cs:0.17, 0.05) {\small $\epsilon = 0.142$};
\node at (axis cs:0.19, 0.14) {\small ${\sqrt{\tfrac{1}{1-\epsilon}}-1}$};

% Panel label
\node[anchor=north west] at (rel axis cs:0,1) {\textbf{(a)}};
%\addplot[black] {sqrt(1/(1 - x)) - 1};
\end{axis}
\node[anchor=east] at (left.west |- yposA) { \textbf{$\tfrac{1}{4\pi}$}};
%%TRY b
% ================= (b) =================
\begin{axis}[
    at={(left.east)},
    anchor=west,
    title={$\kappa(\epsilon)$ necessary to uphold HUP},
    xshift=0.8cm,
    width=8.6cm,
    height=5.2cm,
    axis lines=left,
    xlabel={$\log \epsilon$},
    ylabel={$\log \kappa(\epsilon)$},
    ylabel style={
    at={(axis description cs:0.0,1.05)},
    anchor=south,
    rotate=270
    },
    xmin=-8, xmax=0,
    ymin=0, ymax=8,
    domain=1e-12:exp(-1),
    samples=200,
]
% Exact
\addplot[black, thick]
({log10(x)}, {log10( sqrt(1-x)*(1 + sqrt(1-x)) / (4*pi*x) )});
\node at (axis cs:-4.15,6)
{\small $\log\!\left( \frac{\sqrt{1-\epsilon}\,\bigl(1+\sqrt{1-\epsilon}\bigr)}{4\pi \epsilon} \right)\sim \log\!\left(\frac{1}{2\pi\epsilon}\right)$}; 

\node[anchor=north west] at (rel axis cs:0,1) {\textbf{(b)}};
\end{axis}
\end{tikzpicture}
\caption{(a) GRS as calculated from Equation (9). (b) a log-log plot of Equation (11).}
\end{figure}
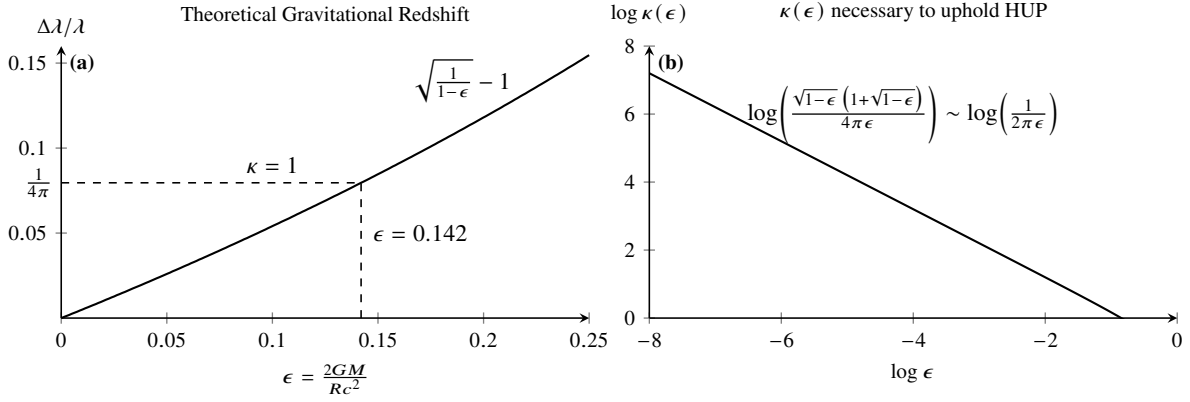
%%%%%%%%FIG 1
We place the ``Heisenberg cut'' at the detector, where photonic quantum states (Fock-space excitations) are converted into measured signals with stochastic intensity $I(\omega)$ and where the Fourier structure enforces $\Delta\omega\,\Delta\tau \gtrsim 1$; the region preceding the cut is therefore the interaction regime in which these quantum states couple to gravity.

Assuming photon excitations propagate through spacetime in accord with GR, the
connection between gravitational redshift and the Heisenberg uncertainty principle takes the form:

\begin{equation}
\frac{1}{2\kappa\pi} \le \frac{\Delta\lambda}{\lambda} = \frac{1}{\sqrt{1-\epsilon}}-1.
\tag{10}
\end{equation}
which provides a lower bound on the fractional wavelength shift due to the HUP and an upper bound due to GR. For large $\epsilon$ values, the wave shifts are in accord with HUP as no special restriction is necessary for $\kappa$ in Equation (10). However, for smaller values of $\epsilon$, $\kappa$ cannot be a mere a constant but must be a monotonously decreasing function of $\epsilon$: \footnote{Because spacetime is highly confined in the vicinity of a black hole, uncertainty relations become relevant. Conversely, in flat spacetime, gravity is a minor perturbation on the Minkowski metric, $g_{\mu\nu}=\eta_{\mu\nu}+h_{\mu\nu}$, permitting a classical description of the photon's wavelength shift.}
\begin{equation}
\kappa(\epsilon)\ge \frac{\sqrt{1-\epsilon}\,\bigl(1+\sqrt{1-\epsilon}\bigr)}{4\pi \epsilon}\sim\frac{1}{2\pi\epsilon}.
\tag{11}
\end{equation}

The ad hoc function $\kappa(\epsilon)$, shown in Fig.~1, is not part of either general relativity or quantum mechanics. Its sole purpose is to preserve consistency with Liouville’s theorem\footnote{Since the spectral width is much larger than the redshift, $\delta\lambda \gg \Delta \lambda$, ~$\kappa(\epsilon)$, keeps the redshift within the uncertainty range in a predictable manner according to GR despite quantum indeterminacy.}: it must be dimensionless, intrinsic to the spacetime manifold, and yet not directly observable. Unlike the coherence phase accumulated by a photon propagating through spacetime (see App.~\ref{app:haag_phase}), 
$\kappa(\epsilon)$ is not restricted to a $2\pi$ range. This behavior suggests a fundamental incompatibility—at macroscopic scales—between GR and QM.
\subsection{A Minimal Model for Photon Propagation in Static Gravity and GRS}
As indicated by Eq.~(2), in a static gravitational field the photon energy, and thus its momentum, is not conserved: 
\begin{equation}
\delta E_{zz'}=h\Delta\nu_{z'z}.
\tag{12}
\end{equation}

Although Eq.~(12) is not manifestly covariant, it is consistent with the covariant formalism of GR. This can be seen as follows. In a static spacetime the condition, $|g_{00}|^{1/2}E_{\text{local}}=\text{const}$, implies
$\partial_z E_{\text{local}}=-(\partial_z |g_{00}|^{1/2})|g_{00}|^{-1/2}E_{\text{local}}$. In the weak--field limit,
\[
g_{00}(z)\simeq -\left(1+\frac{2\Phi(z)}{c^{2}}\right),
\qquad
|g_{00}|^{1/2}\simeq 1+\frac{\Phi(z)}{c^{2}},
\]
Leading to, 
$\partial_z|g_{00}|^{1/2}\simeq (1/c^{2})\,\partial_z\Phi$. Substituting and expanding to first order in $\Phi/c^{2}$,
\[
\frac{1}{1+\Phi/c^{2}} \simeq 1,
\] we obtain,
\[
\partial_z E_{\text{local}}
\simeq -\frac{1}{c^{2}}
\frac{\partial_z\Phi}{1+\Phi/c^{2}}\,E_{\text{local}}
\simeq -\frac{1}{c^{2}}\,\partial_z\Phi\,E_{\text{local}}.
\]
For a photon, $p_z=E_{\text{local}}/c$, giving
\[
\partial_z p_z\simeq -\frac{E_{\text{local}}}{c^{3}}\,\partial_z\Phi,
\]
showing that spatial changes in the photon momentum are driven directly by the gradient of the gravitational potential~$\Phi(z)$.
\\
\\
Also in QED, where local field operators satisfy the Heisenberg equation
$\partial_t\hat{\mathcal{O}}(x)=\tfrac{i}{\hbar}[\hat{H},\hat{\mathcal{O}}(x)]$
with $\hat{H}=\int d^3x\,\mathcal{H}_{\mathrm{QED}}(x)$, the momentum is not conserved in a gravitational gradient, since
$\partial_t\langle\hat{\mathcal{P}}_z\rangle = -\langle\partial_z\mathcal{H}_{\mathrm{QED}}\rangle \neq 0$ (see Appendix~\ref{app:QED}).
\\
\\
\par
We use the gravitational redshift to construct a semiclassical model in which photon propagation in a static gravitational field is represented as a succession of local interactions, each producing an infinitesimal wavelength shift. In contrast to Haag’s model of $\alpha$-particle ionization in a cloud chamber~\cite{Haag}, our approach assumes the local realism of the spacetime metric, which serves as the measurement apparatus. Each interaction occurs with unit probability, $Prob(z)=1$, and therefore induces a definite change in the photon’s momentum through its wavelength shift. A sequence of such interactions accumulates into an observable outcome registered by a detector, the only stage at which the ``Heisenberg cut'' is required to encounter the uncertainty principle. We show that this construction leads to a paradox.
\\
\\
\par
For a photon traveling the distance $\Delta z = z'-z$, in a static gravitational field, the change in momentum from $p_{z}=h/\lambda_{z}$, to, $p_{z'}=h/\lambda_{z'} = h/(\lambda_{z}+\Delta\lambda_{zz'})$, at $z'$, is:
\begin{equation}
\Delta p_{zz'} = h\left(\frac{1}{\lambda_{z'}}-\frac{1}{\lambda_{z}}\right)
= -h\frac{\Delta\lambda_{zz'}}{\lambda_{z}(\lambda_{z}+\Delta\lambda_{zz'})}.
\tag{13}
\end{equation}
Dividing both the numerator and denominator by $\lambda_{z}$ we obtain:
\begin{equation}
\Delta p_{zz'} =
\left(\frac{-h}{\lambda_{z}}\right)
\left(
\frac{\Delta\lambda_{zz'}/\lambda_{z}}
{1+\Delta\lambda_{zz'}/\lambda_{z}}
\right).
\tag{14}
\end{equation}
For $\Delta\lambda_{zz'}/\lambda_{z}\ll 1$, and to first order, we have:
\begin{equation}
\Delta p_{zz'} \cong
\left(\frac{-h}{\lambda_{z}}\right)
\left(\frac{\Delta\lambda_{zz'}}{\lambda_{z}}\right)
=
\frac{-h\Delta\lambda_{zz'}}{\lambda_{z}^{2}}.
\tag{15}
\end{equation}
The momentum-position uncertainty relation follows Equations (3) but at the same time cannot be larger than $\Delta p_{zz'}\times(z'-z)$. Here again we cast the distance traveled in term of the wavelength as $\Delta z = z'-z = \varkappa\lambda_{z}$, to arrive at the following inequality:
\begin{equation}
\frac{1}{2\pi}\le \frac{\varkappa\Delta\lambda_{zz'}}{\lambda_{z}}.
\tag{16}
\end{equation}
This inequality is strongly dependent on the constant, $\varkappa$, but since the relative wavelength shift is very small, $\Delta\lambda_{zz'}/\lambda_{z}\approx 10^{-15}$, for Equation (16) to hold true, $\varkappa$ has to be of the order of $\varkappa\sim O(10^{15})$, which is highly unlikely. This result sharpens the paradox suggested above (Section II.A). At each position during its propagation through a static gravitational field, while losing energy via GRS, the photon has to have real simultaneous continuous variables, momentum and position, while at the same time it must succumb to the uncertainty principle which states the opposite.

\section{HUP and Entanglement}

In the sections above, evidence for tension between GRS and HUP was shown both from the empirical and theoretical aspects. This, albeit conceptual tension, suggests that already at the weak gravitational field, gravitational redshift may be incompatible with the Heisenberg uncertainty principle. If the Heisenberg assertion that momentum and position cannot assume simultaneously physical realities is false, then the picture painted by general relativity would explain GRS, even for a quantum object such as a photon, rather classically. The very fact that the uncertainty relations as per Equations (4,5), is shedding the Planck's constant, $h$, encapsulates succinctly this message. Namely, the uncertainty relations lose their quantum context and wholly conform to a classical picture\footnote{A brief discussion is given in Appendix~\ref{app:haag_phase}, in connection with the Tokyo Skytree measurement of GRS~\cite{Takamoto}, which compares stationary optical clocks and resolves the gravitational redshift while effectively circumventing any role for the HUP.}.

Photonic entanglement deems quantum mechanics a nonlocal theory, in contradiction with general relativity and the equivalence principle. This is best described by the EPR paradox \cite{EPR1935}. Entangled photons that undergo gravitational redshift would entangle spacetime itself, thereby turning it into a quantum entity. So far, as has been observed in numerous tests of general relativity, spacetime adheres to its local character. Extending the argument of local realism from single quanta photons to entangled photons would mean that the entangled pair violates the uncertainty principle with respect to the independent reality of their positions and momenta in spacetime. No ``steering''\footnote{Schr\"odinger introduced the concept of entanglement or ``steering'' to express the nonlocal effect of part of a system on another part of it, in what seems to be an instantaneous action at a distance.} information has to travel through spacetime faster than the speed of light in order to maintain the correlation between the members of the pair. The initial correlation would be altered by the local interaction of each photon of the pair with spacetime. As no measurement is performed the ``collapse of the wavefunction'' is avoided altogether.

At this point we note that the observed entanglement in the polarization of photons does not cause any tension between quantum mechanics and general relativity, as no spin-dependent effects are expected in a weak gravitational field. This is because intrinsic spin is equivalent to ordinary angular momentum. If this were not the case, the equivalence principle would be violated, however, thus far all empirical evidence support it \cite{Wei-Tou}. The entanglement in helicity (polarization) states is therefore a feature of quantum mechanics alone. One can invoke the second law of thermodynamics in providing contextual arguments for why quantum entanglement in helicity states does not pose a fundamental difficulty for general relativity while GRS does. Since the entangled pair remains a closed system with respect to their coupled helicity states, and no energy is dispersed through spin--gravity interaction, the system's entropic information remains unchanged \cite{Cao}. This, however, does not hold for the photons' momenta, as indicated in Equation (12). The energy dissipated when a photon undergoes gravitational redshift necessarily leads to an increase in entropy.

Assume a pair of entangled photons, one traveling a vertical distance $z$ above Earth's surface and undergoing gravitational redshift. Its entangled partner remains at the surface, ($z=0$), so its wavelength does not shift. How does GRS affect the entangled partner that doesn't experience it directly?

To this end we propose to test the effect with a thought experiment utilizing a set of EPR entangled beams on Earth's surface. Entanglement-based EPR arguments presuppose a single global time parameter $t$ with respect to joint observables such as:
\[
\widehat{X}_{-} = \widehat{x}_{A}(t)-\widehat{x}_{B}(t) ~ \text{and} ~ 
\widehat{P}_{+} = \widehat{p}_{A}(t)+\widehat{p}_{B}(t), 
\]
which are defined as simultaneous measurements on spatially separated subsystems. In a curved spacetime, by contrast, each subsystem follows a worldline, $x_{j}^{\mu}(\tau_{j})$, with its own proper time, $\tau_{j}$, related to any chosen coordinate time $t$ by:
\[
d\tau_{j} = \sqrt{-g_{\mu\nu}(x_{j})\dot{x}_{j}^{\mu}\dot{x}_{j}^{\nu}}\,dt,
\]
which in the weak-field limit reduces to, $d\tau_{j}\simeq \left(1+\Phi_{j}/c^{2}\right)dt$.
The local mode operators evolve as:
\[
\widehat{a}_{j}(\tau_{j}) = \widehat{a}_{j}(0)e^{-i\omega_{j}\tau_{j}},
\qquad
\widehat{a}^{\dagger}_{j}(\tau_{j}) = \widehat{a}^{\dagger}_{j}(0)e^{-i\omega_{j}\tau_{j}},
\]
so that expressing both subsystems in a single time parameter $t$ induces unequal effective frequencies:
\[
\omega_{j}^{\mathrm{eff}}=\omega_{j}\left(1+\Phi_{j}/c^{2}\right),
\]
and, at the level of the canonical quadratures \cite{Furusawa}:
\[
\widehat{x}_{j}=\frac{1}{\sqrt{2}}\left(\widehat{a}_{j}+\widehat{a}_{j}^{\dagger}\right),
\qquad
\widehat{p}_{j}=\frac{1}{\sqrt{2}}\left(\widehat{a}_{j}-\widehat{a}_{j}^{\dagger}\right),
\]
a nontrivial symplectic map:
\[
(\widehat{x}_{j},\widehat{p}_{j})\mapsto (\sqrt{\zeta_{j}}\,\widehat{x}_{j},\,\widehat{p}_{j}/\sqrt{\zeta_{j}}) ~ \text{with} ~ \zeta_{j}\simeq \left(1+\Phi_{j}/c^{2}\right).
\]

The global EPR observables $\widehat{X}_{-}$ and $\widehat{P}_{+}$ become explicitly gravity-dependent constructs whose canonical form is not invariant under changes of gravitational potential, even though the underlying bipartite state should remain entangled\footnote{This issue has, in fact, already manifested in the discussion of a single M\"ossbauer wavepacket (Sec.~II.A).}. Thus, the mathematical structure of general relativity---through the replacement of a universal time $t$ by different proper times $\tau_{A},\tau_{B}$---undermines the very notion of globally defined, simultaneously measurable EPR observables, exposing a tension between the EPR state and dilation of time in a weak gravitational field. Moreover, the entanglement of the $\lvert \mathrm{EPR}\rangle$ state eliminates the uncertainty of the combined variables, $\widehat{X}_{-}$ and $\widehat{P}_{+}$. Specifically:
\[
[\widehat{x}_{A}-\widehat{x}_{B},\,\widehat{p}_{A}+\widehat{p}_{B}]
=
[\widehat{x}_{A},\widehat{p}_{A}]
+
[\widehat{x}_{A},\widehat{p}_{B}]
-
[\widehat{x}_{B},\widehat{p}_{A}]
-
[\widehat{x}_{B},\widehat{p}_{B}]
=0.
\]
This can be described with quantum mechanical expectation values as:
\[
\left\langle \left[\Delta\left(\widehat{x}_{A}-\widehat{x}_{B}\right)\right]^{2} \right\rangle
+
\left\langle \left[\Delta\left(\widehat{p}_{A}+\widehat{p}_{B}\right)\right]^{2} \right\rangle
=0.
\]
with the vanishing sum reflecting the choice of a simultaneous eigenstate of the two compatible observables \footnote{Since $(\widehat{x}_{A}-\widehat{x}_{B})|\mathrm{EPR}\rangle=0$ and $(\widehat{p}_{A}+\widehat{p}_{B})|\mathrm{EPR}\rangle=0$, it follows that $\Delta(\widehat{x}_{A}-\widehat{x}_{B})=\Delta(\widehat{p}_{A}+\widehat{p}_{B})=0$, and therefore $\langle[\Delta(\widehat{x}_{A}-\widehat{x}_{B})]^{2}\rangle+\langle[\Delta(\widehat{p}_{A}+\widehat{p}_{B})]^{2}\rangle=0$.}. 

The meaning of this, as EPR articulated in their paper, is that nonlocal correlations allow these joint observables to evade the usual Heisenberg uncertainty constraints that apply to individual subsystems. In other words, if entanglement makes it possible to assign definite values to these combined variables without violating the commutation relations, then---according to the EPR reasoning---those variables must correspond to elements of physical reality.

The entanglement of the EPR state $\lvert \mathrm{EPR}\rangle$ can be described using either continuous or discrete bases, but it is not normalizable and therefore represents an unphysical quantum state. In practice, one uses the two-mode squeezed vacuum, denoted $\lvert \mathrm{EPR}^{\ast}\rangle$, which approaches the idealized $\lvert \mathrm{EPR}\rangle$ state asymptotically, reproducing its correlations arbitrarily well in the limit \cite{Furusawa}.

\subsection{Proposed Experiment}

As gravitational redshift offers the stage on which quantum mechanics is pitted against general relativity we propose to use photonic entanglement as an experimental testbed of local realism per EPR. In quantum mechanics a pair of entangled photons is viewed as a single system which maintains its correlation as long as it is not subject to collapse due to measurement. Even if the system undergoes (quantum) decoherence \cite{Schlosshauer} due to gravity \footnote{Induced gravitational quantum phase shifts were first observed in neutron‑interference experiments by Colella, Overhauser, and Werner \cite{COW}. Likewise, massless photons in curved spacetime acquire a phase, $\phi = \frac{1}{\hbar} \int p_\mu \, dx^\mu$, determined by the underlying geometry \cite{Stodolsky}. Decoherence does not alter the phase itself; rather, it governs whether quantum states can interfere or become distinguishable. Within this viewpoint, the system could be regarded as becoming entangled with quantum degrees of freedom attributed to spacetime, leading to the suppression of interference and the loss of off‑diagonal density‑matrix elements when those degrees of freedom are traced out. However, any such attribution remains speculative: the existence and nature of these putative degrees of freedom would depend on a theory of quantum gravity that is not currently established. Consequently, this interpretation should be regarded with appropriate caution.} the lifetime of the correlation between the photons of the pair can be estimated by the coherence length:
\[
L_{c}=\frac{\lambda}{\left(\delta\lambda/\lambda\right)},
\qquad
\tau_{c}\approx \frac{L_{c}}{c}.
\]
Should the entangled pair obey the uncertainty relation, then, within the time estimated above, the correlation between the members of the pair should stay unperturbed. Hence, by testing if entanglement withstands gravitational redshift, or within what time it breaks down, we should be able to test the effect of a gravitational field on a quantum system. More precisely, we seek to determine whether the entanglement of the pair enforces the conservation of the system's initial energy after one of its members has traveled a height $z$:
\[
\Delta E_{\text{pair}}
= \hbar(\omega^{(A)}_{h=0}
- \omega^{(B)}_{h=0})
\;\overset{?}{=}\;
\hbar (\omega^{\text{(A)}}_{h=0}
- \omega^{\text{(B)}}_{h=z}).
\]
If the initial energy of the entangled EPR state were shown to remain conserved after one constituent has been displaced to a height $z$, such conservation would necessarily be attributable solely to the underlying entanglement structure. This result would, in turn, provide evidence that quantum‑mechanical degrees of freedom play an essential role in gravitational contexts and therefore must be incorporated into the theoretical framework of general relativity.

As depicted schematically in figure 2, Alice would be observing at ground level the interference pattern that arises between the laser beams A and A', both with the same wavelengths.

While A' is entangled with A'', A is entangled with B. Beam B undergoes redshift, while all the A beams do not. Bob's measurement at a height $z$ (far enough from ground level) establishes that beam B is redshifted. The difference between the energy of Bob's beam to that of Alice's is, $\Delta E = \frac{h\nu}{c^{2}}\Delta U_{G}$, where, $U_{G}$, is the gravitational potential.

\begin{figure}[t]
\centering
\includegraphics[width=0.78\linewidth]{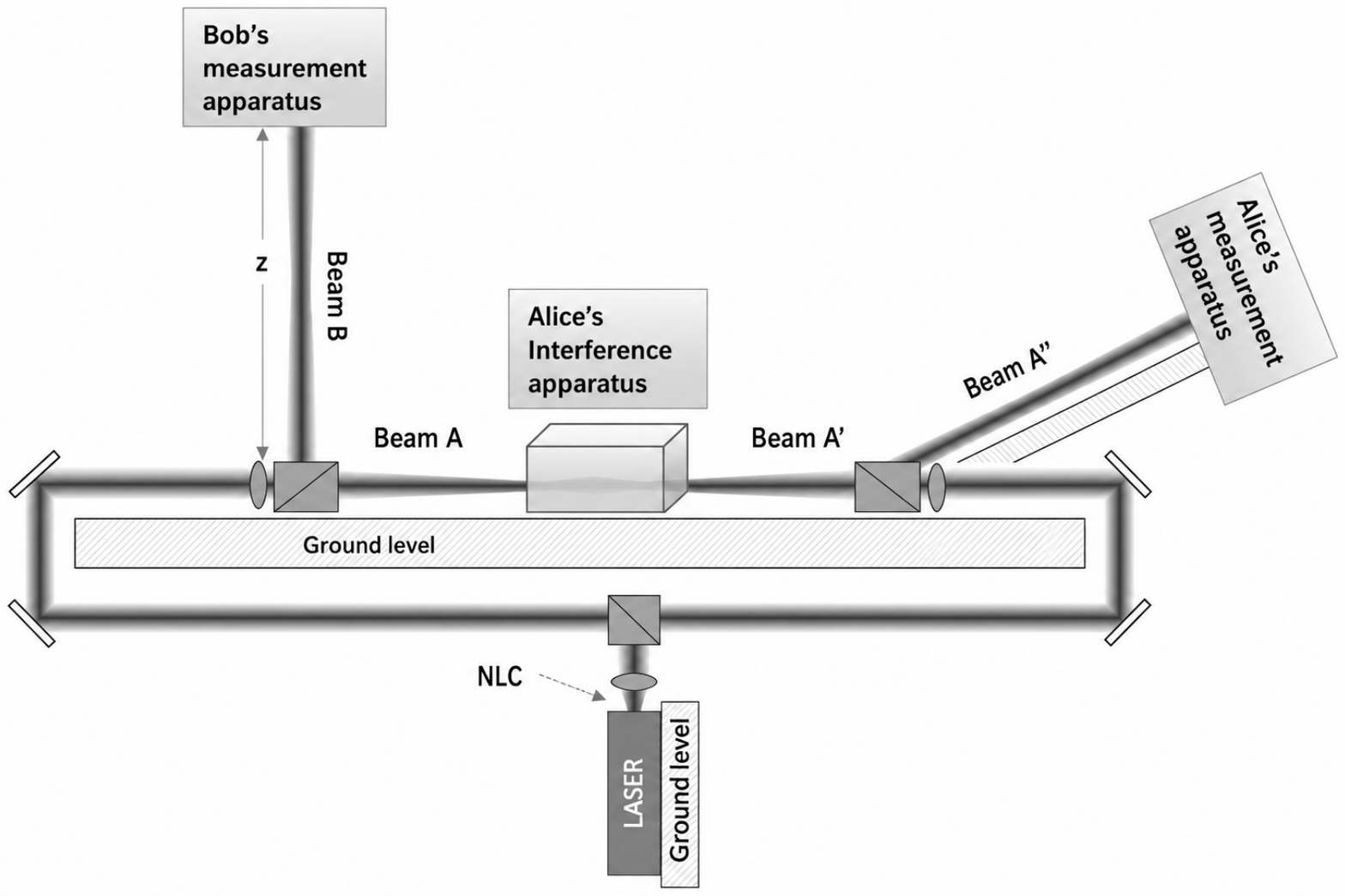}
\caption{A general scheme of a proposed Mach--Zehnder interferometer for the experimental setup.}
\end{figure}

Alice's measurement of beam A'' serves as an indirect verification of the wavelength of beam A', confirming its properties as initially generated. If beam B, which undergoes gravitational redshift, influences the wavelength of beam A due to entanglement, then the interference pattern formed between beams A and A' will be modified. This would imply a nonlocal action at a distance---manifested as an observed change of the interference pattern. Thus, within a short time frame, Alice should be able to detect changes in the interference pattern between A and A', induced by the gravitational redshift experienced by the entangled partner beam B.

The experiment could exhibit the following possible outcomes:
\begin{enumerate}[label=\arabic*.]
\item The interference pattern changes only after Bob performs his redshift measurement.
\item Bob doesn't measure redshift on his beam but the interference pattern changes on Alice's end within the above estimated lifetime.
\item The interference pattern changes continuously from the onset of the experiment reaching a maximum around the time that it takes the beam to reach Bob's location at $z$, after $z/c$ sec.
\item The interference pattern doesn't change whether Bob performs his measurement or not.
\end{enumerate}

While the first three outcomes indicate that entanglement influences spacetime and may therefore require incorporation into general relativity, the fourth outcome instead suggests that the photon interaction with spacetime is fundamentally local and that GR does not require quantum degrees of freedom at all.

In addition, while Alice monitors the interference pattern on her end, spacetime may effectively function as a kind of ``mindless Wigner's friend'' acting on beam B. That is, as beam B being redshifted while traveling upwards, entanglement may lead to a gradual degradation of the initially prepared interference pattern---or else the pattern may remain unaffected.

By contrast, when Bob measures the redshifted wavelength on his apparatus, he effectively becomes Alice's ``mindful Wigner's friend,'' as his conscious observation directly interacts with the quantum system \cite{Wigner}. If Bob's measurement induces an abrupt change in the interference pattern, then his observation constitutes an active, conscious intervention in the entangled state.

 Because the beams are polarization entangled, Bell‑inequality tests can be performed concurrently with the continuous‑variable entanglement. A polarized, entangled‑beam arrangement analogous to Ref.~\cite{COW} would offer the same capability. Appendix~\ref{app:bell_phase} sketches the underlying theory.
\section*{Summary}
We highlight a fundamental tension between the Heisenberg uncertainty principle and gravitational redshift, giving rise to the following paradox: how can gravitational redshift occur when a photon's conjugate variables---momentum and position---cannot be simultaneously well defined in a manner compatible with local realism as articulated by EPR? To examine this issue, we propose a thought experiment employing continuous-variable quantum entanglement---an approach that has not yet been experimentally realized---in the presence of a weak gravitational field. The outcome of such an experiment may extend, and potentially empirically establish, the known incompatibility between quantum mechanics and general relativity into the weak-gravity regime.

%Because the beams are polarization entangled, the setup admits Bell‑inequality measurements concurrent with the continuous‑variable entanglement at both ends, thereby enabling a direct comparison between the two forms of entanglement.

\section*{Acknowledgements}

Asher Klatchko gratefully acknowledges David Griffiths, Emeritus Professor of Physics at Reed College, Robert Hamilton of George Fox University, Sandu Popescu of Bristol University and Daniel Rub\'en Zerzi\'on of~DIPC San Sebasti\'an for insightful discussions and valuable comments. We also wish to acknowledge productive comments by Germano D'Abramo from Ministero dell'Istruzione, Università e Ricerca.

\appendix

\section{Inferred Coherence}
\label{app:haag_phase}
Coherence is not an observable but a property inferred from interference and correlation measurements. For a state with density matrix $\rho$, the off-diagonal elements encode coherence and determine interference effects. A measurable proxy is the fringe visibility,  $V=(I_{\max}-I_{\min})/(I_{\max}+I_{\min})$, with $0\le V\le1$. Coherence appears in two-point functions $G(x,y)=\langle\phi(x)\phi(y)\rangle$, whose non-factorization signals quantum correlations. In algebraic QFT, a state $\omega$ is coherent when $\omega(AB)\neq\omega(A)\omega(B)$ for local observables $A$ and $B$. Operationally, coherence is the deviation from statistical independence of observables\cite{Schlosshauer}.
\\
In relativistic wave mechanics and geometric optics, the phase of a wave is described by the eikonal function $S(x)$, whose gradient defines the wave covector:
\begin{equation}
k_\mu = \partial_\mu S .
\end{equation}

The accumulated phase along a spacetime trajectory is therefore
\begin{equation}
\phi = \int k_\mu \, dx^\mu .
\end{equation}

For electromagnetic radiation in vacuum, the wave four-vector is null:
\begin{equation}
k^\mu k_\mu = 0 .
\end{equation}
The quantity $k_\mu dx^\mu$ is a Lorentz scalar, and therefore the phase
$\phi$ is invariant under coordinate transformations.
\\
In the Pound--Rebka experiment, coherence is detected through resonance absorption, with the rate proportional to $|\langle A_a A_e\rangle|^2$. The exchange between emitter and absorber is governed by the invariant phase $\phi$, and maximal absorption requires $\Delta\phi_{\rm total}=0$. The emitter and absorber contribute local phases $\phi_{e,a}=\omega_{e,a}\tau_{e,a}$, with the absorber frequency shifted by $\omega_a\to\omega_a(1+v/c)$. In a stationary spacetime the measured frequency is $\omega=-u^\mu k_\mu$, giving the phase-matching condition
\[
\frac{E}{\sqrt{g_{00}(x_e)}}=
\frac{E}{\sqrt{g_{00}(x_a)}}\!\left(1+\frac{v}{c}\right).
\]
Using $g_{00}(z)\simeq 1+2gz/c^2$ yields
\[
\frac{v}{c}=\frac{gh}{c^2}.
\]
Thus the experiment enforces the invariant equality
\[
(u^\mu k_\mu)_e=(u^\mu k_\mu)_a ,
\]
with the absorber’s Doppler motion compensating the gravitational redshift.
\\
\\
%%%%%%
Note that this is not the case in the Tokyo Skytree experiment \cite{Takamoto}, where no Doppler tuning was applied to cancel the gravitational frequency offset between clocks. Instead, the clocks were continuously compared via an optical link \cite{Takamoto1}, and the directly accessible observable was the fractional frequency difference $\Delta\nu/\nu$. This frequency offset corresponds to a nonzero accumulated relative phase, although only the frequency (i.e. the phase evolution rate) is directly measured\footnote{The measured frequency is the time derivative of the accumulated phase, $\omega = d\phi/dt$, so clock-comparison experiments effectively measure differences in phase evolution rates rather than absolute phase offsets.}.
\\
Locally, the clock frequency is given by $\omega = -u^\mu k_\mu$, yielding
\[
\Delta\omega = (u^\mu k_\mu)_{\rm top} - (u^\mu k_\mu)_{\rm base},
\]
and the corresponding phase accumulation
\[
\Delta\phi = \int \Delta\omega\, d\tau,
\]
defined with respect to a common spacetime foliation. In this framework, however, the reported observable is fundamentally a frequency difference inferred from the time derivative of locally accumulated phases, rather than a direct measurement of an inter-clock phase difference $\phi_{\rm top} - \phi_{\rm base}$.
\\
Further note that the instability of the clock-comparison signal, i.e. the statistical uncertainty in estimating $\Delta\nu/\nu$, is ultimately limited by quantum projection noise (QPN), yielding a QPN-limited fractional frequency instability at the level $\Delta\nu_{\rm QPN}/\nu_0 \sim 10^{-18}$. For the Sr-based optical lattice clocks used in the experiment, this regime is reached by suppressing systematic shifts such as blackbody radiation and higher-order light shifts. These statistical uncertainties are several orders of magnitude smaller than the observed gravitational redshift,
\[
\frac{\Delta\nu}{\nu} \simeq (1+\alpha)\frac{g\Delta z}{c^2} \approx 4.9 \times 10^{-14}
\quad \text{for } \quad \Delta z = 452.6\,\mathrm{m},
\]
with a deviation parameter $\alpha = (1.4 \pm 9.1)\times 10^{-5}$ relative to general relativity ($\alpha = 0$), thereby improving upon the precision of the Pound--Rebka experiment ($\alpha \sim 10^{-2}$) by approximately two to three orders of magnitude.
\section{Field Momentum Density and Ehrenfest theorem in QED}
\label{app:QED}
For photons the field momentum density follows from the $0z$ component of the
stress--energy tensor,
\[
\hat{\mathcal{P}}_z \equiv \hat{T}^{0z}
= \hat{F}^{0\lambda}\hat{F}^{z}{}_{\lambda}
= \hat{E}_i \hat{F}_{zi},
\]
and with $F^{0i}=E^i$ and $F_{zi}=\epsilon_{zik}B_k$ this becomes
\[
\hat{E}_i \hat{F}_{zi} = (\hat{\mathbf{E}}\times\hat{\mathbf{B}})_z,
\]
the standard Poynting momentum density.
The Heisenberg equation,
\[
\partial_t \hat{\mathcal{O}}(x)
= (i/\hbar)[\hat{H},\hat{\mathcal{O}}(x)],
\]
with $\hat{H}=\int d^3x\,\mathcal{H}_{\mathrm{QED}}(x)$,
together with the canonical commutator
\[
[\hat{A}^i(x),\hat{E}^j(y)]
= i\hbar\,\delta^{ij}\delta^3(x-y),
\]
gives
\[
\partial_t \hat{A}_i = (i/\hbar)[\hat{H},\hat{A}_i],
\qquad
\partial_t \hat{E}_i = (i/\hbar)[\hat{H},\hat{E}_i],
\]
which reproduce the operator Maxwell equations
\[
\partial_\mu \hat{F}^{\mu\nu}=0
\]
for free photons.
\\
The relation
\[
\nabla_\mu T^{\mu z} = -F^{z\mu} j_\mu
\]
expresses how the field's $z$-momentum changes in the presence of charges.
For photons $j_\mu=0$, so $\nabla_\mu T^{\mu z}=0$ expresses local momentum
conservation, but the covariant divergence contains metric--gradient terms,
meaning that a spatially varying metric allows $\partial_\mu T^{\mu z}\neq 0$
even though $\nabla_\mu T^{\mu z}=0$.
In static gravity this means that the field spatial gradients alter the photon
momentum, so $\partial_0 T^{0z}\neq 0$ and $\hat{\mathcal{P}}_z$ is not
conserved. The conservation law
\[
\partial_\mu \hat{T}^{\mu\nu}(x)=0
\]
then gives
\[
\partial_t \hat{T}^{0\nu}(x)
= -\partial_i \hat{T}^{i\nu}(x),
\]
and taking expectation values yields
\[
\partial_t \langle \hat{T}^{0\nu}(x) \rangle
= -\partial_i \langle \hat{T}^{i\nu}(x) \rangle.
\]
For $\hat{\mathcal{P}}_z=\hat{T}^{0z}$ this becomes
\[
\partial_t \langle \hat{\mathcal{P}}_z(x) \rangle
= -\partial_i \langle \hat{T}^{iz}(x) \rangle,
\]
showing explicitly that the photon momentum is not conserved in a spatially varying gravitational field.

\section{Bell type experiment with a gravitational induced phase shift}
\label{app:bell_phase}
This appendix contrasts gravitationally induced phase shifts in polarization entanglement with the field‑dependent energy effects in continuous‑variable entanglement.

Consider a polarization-entangled photon pair prepared in the Bell state
\begin{equation}
|\Psi_0\rangle = \frac{1}{\sqrt{2}}\left(|H\rangle_A |H\rangle_B + |V\rangle_A |V\rangle_B\right).
\end{equation}
Photon $A$ propagates along a horizontal reference path, while photon $B$ follows a trajectory $\gamma$ in a weak, static gravitational field. In the weak-field limit, the metric is
\begin{equation}
ds^2 = -\left(1 + \frac{2gz}{c^2}\right)c^2 dt^2 + d\mathbf{x}^2.
\end{equation}
The phase accumulated by a photon along $\gamma$ is
\begin{equation}
\phi = \int_\gamma k_\mu dx^\mu,
\end{equation}
where $k^\mu$ is the null wavevector. In a stationary spacetime, the conserved quantity
\begin{equation}
\omega = -k_\mu \xi^\mu
\end{equation}
(with $\xi^\mu = (\partial_t)^\mu$) defines the photon frequency.
To leading order in $gz/c^2$, the phase difference between two paths reduces to
\begin{equation}
\phi = \omega \Delta \tau,
\end{equation}
where the effective proper-time difference is
\begin{equation}
\Delta \tau = \frac{1}{c^2} \int g z(t)\, dt.
\end{equation}
Thus,
\begin{equation}
\phi = \frac{\omega}{c^2} \int g z(t)\, dt.
\end{equation}
After propagation, the entangled state becomes
\begin{equation}
|\Psi\rangle = \frac{1}{\sqrt{2}}\left(|H\rangle_A |H\rangle_B + e^{i\phi} |V\rangle_A |V\rangle_B\right).
\end{equation}
This phase is observable in joint polarization measurements. In the basis
\begin{equation}
|\pm\rangle = \frac{1}{\sqrt{2}}(|H\rangle \pm |V\rangle),
\end{equation}
the coincidence probability reads
\begin{equation}
P_{++} = P_{--} = \frac{1}{2}\left(1 + \cos \phi\right) ~\text{and}
\end{equation}
\begin{equation}
P_{+-} = P_{-+} = \frac{1}{2}\left(1 - \cos \phi\right).
\end{equation}
As Eq.~(C7) shows, the gravitational field induces a Bell‑detectable relative phase that, depending on the measurement basis, can at most obscure—but not eliminate—the entanglement, whereas the redshift in Eq.~(1) can destroy the underlying momentum‑entanglement structure altogether.
%%%%%tikz figure
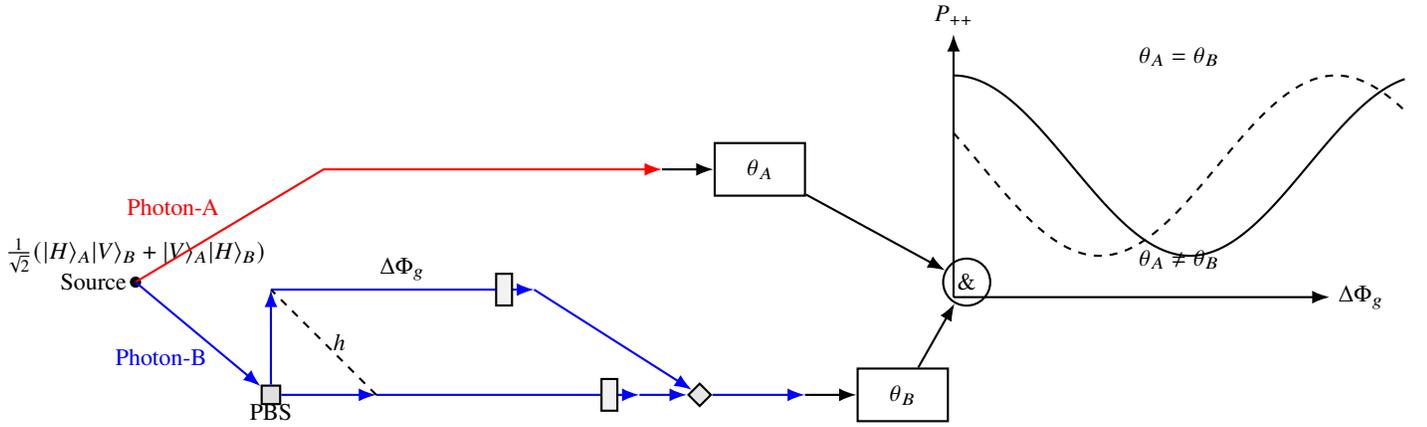
\begin{figure}[t]
\centering
\usetikzlibrary{arrows.meta,calc,decorations.pathmorphing,shapes,positioning}

    \begin{tikzpicture}[>=Latex, thick]

    % =========================
    % PANEL (a): SETUP
    % =========================

    \begin{scope}[shift={(0.375,0.2)}]

    \tikzset{
    photonA/.style={red},
    photonB/.style={blue},
    pbs/.style={draw, fill=gray!25, minimum size=7pt},
    bs/.style={draw, fill=gray!20, minimum size=6pt, rotate=45},
    hwp/.style={draw, fill=gray!10, minimum width=6pt, minimum height=12pt},
    det/.style={draw, minimum width=1.2cm, minimum height=0.7cm}
    }

    % Source
    \coordinate (S) at (0,0);
    \node[circle, fill=black, inner sep=1.5pt] at (S) {};
    \node[left] at (S) {\small Source};

    \node[above] at (S) {\small 
    $\frac{1}{\sqrt{2}}(|H\rangle_A|V\rangle_B + |V\rangle_A|H\rangle_B)$};

    % Photon A
    \draw[photonA,->] (S) -- ++(2.5,1.5) -- ++(4.5,0) coordinate (Aout);
    \node[text=red] at (0.5,1) {\small Photon-A};
    \node[det] (DA) at ($(Aout)+(1.3,0)$) {};
    \node at (DA) {\small $\theta_A$};
    \draw[->] (Aout) -- (DA);

    % Photon B
    \coordinate (Bstart) at ($(S)+(1.8,-1.5)$);
    \node[pbs] (PBS) at (Bstart) {};
    \node[below] at (PBS) {\small PBS};
    \node[text=blue] at (0.34,-1) {\small Photon-B};
    \draw[photonB,->] (S) -- (PBS);

    % Split
    \coordinate (L1) at ($(PBS)+(1.4,0)$);
    \coordinate (U1) at ($(PBS)+(0,1.4)$);

    \draw[photonB,->] (PBS) -- (L1);
    \draw[photonB,->] (PBS) -- (U1);

    % Arms
    \coordinate (L2) at ($(L1)+(3.5,0)$);
    \coordinate (U2) at ($(U1)+(3.5,0)$);

    \draw[photonB,->] (L1) -- (L2);
    \draw[photonB,->] (U1) -- (U2);

    % Height + phase
    \draw[dashed] (L1) -- (U1);
    \node[right] at ($(L1)!0.5!(U1)$) {\small $h$};

    \node[above] at ($(U1)!0.5!(U2)$) {\small $\Delta\phi_g$};

    % HWP (eraser)
    \node[hwp] at ($(L2)+(-0.4,0)$) {};
    \node[hwp] at ($(U2)+(-0.4,0)$) {};

    % BS
    \node[bs] (BS2) at ($(L2)+(0.8,0)$) {};
    \draw[photonB,->] (L2) -- (BS2);
    \draw[photonB,->] (U2) -- (BS2);

    \coordinate (Bout) at ($(BS2)+(1.4,0)$);
    \draw[photonB,->] (BS2) -- (Bout);

    % Detector B
    \node[det] (DB) at ($(Bout)+(1.3,0)$) {};
    \node at (DB) {\small $\theta_B$};
    \draw[->] (Bout) -- (DB);

    % Coincidence
    \node[circle, draw] (C) at ($(DA)!0.5!(DB)+(1.8,0)$) {\small $\&$};
    \draw[->] (DA) -- (C);
    \draw[->] (DB) -- (C);

    \end{scope}

    % =========================
    % PANEL (b): FRINGES
    % =========================

    \begin{scope}[shift={(11.25,0)}]

    % Axes
    \draw[->] (0,0) -- (5,0) node[right] {\small $\Delta\phi_g$};
    \draw[->] (0,0) -- (0,3.5) node[above] {\small $P_{++}$};

    % Sine curves
    \draw (0,1.75)
      plot[domain=0:6,samples=100]
      (\x,{1.75 + 1.2*cos(deg(\x))});

    \draw[dashed] (0,1.75)
      plot[domain=0:6,samples=100]
      (\x,{1.75 + 1.2*cos(deg(\x+1.2))});

    % Labels
    \node at (3,3.2) {\small $\theta_A=\theta_B$};
    \node at (3,0.5) {\small $\theta_A \neq \theta_B$};

    \end{scope}
    \end{tikzpicture}

    \caption{Schematic of an entangled--photon interferometer in which
    Photon~A follows a horizontal reference path while Photon~B traverses
    a vertical interferometer and acquires a gravitationally induced
    phase~$\phi$. The phase shift is read out through
    polarization--correlation measurements in coincidence.}
    \label{fig:entangled_interferometer}

\end{figure}

%%%%%tikz figure
\newpage
\section*{References}
%\begin{enumerate}[label={[\arabic*]},leftmargin=2.5em]
\bibliographystyle{apsrev4-2}

\end{document}